\documentclass[prb,twocolumn,amscd,amsmath,amssymb,verbatim]{revtex4}

\usepackage{graphicx}
\usepackage[dvips]{hyperref}
\usepackage[TS1,OT1,T1]{fontenc}

\begin{document}
\title{Kondo behavior, ferromagnetic correlations, and crystal fields in the heavy Fermion compounds Ce$_3$X ( X=In, Sn)}

\author{C. H. Wang$^{1,2}$, J. M. Lawrence$^1$, A. D. Christianson$^3$, E. A. Goremychkin$^{4}$,
V. R. Fanelli$^2$, K. Gofryk$^2$, E. D. Bauer$^2$, F. Ronning$^2$,
J. D. Thompson$^2$, N. R. de Souza$^{4,5}$, A. I. Kolesnikov$^{3}$,
K. C. Littrell$^3$}
\affiliation{$^1$University of California, Irvine, California 92697, USA\\
  $^2$Los Alamos National Laboratory, Los Alamos, NM 87545, USA\\
  $^3$Neutron Scattering Sciences Division, Oak Ridge National Laboratory, Oak Ridge, TN, 37831, USA\\
  $^4$Argonne National Laboratory, Argonne, IL 60439, USA\\
  $^5$Australian Nuclear Science and Technology Organisation, Lucas Heights, NSW 2234, Australia}

\date{\today}

\begin{abstract}

We report measurements of inelastic neutron scattering, magnetic
susceptibility, magnetization, and the magnetic field dependence of
the specific heat for the heavy Fermion compounds Ce$_3$In and
Ce$_3$Sn. The neutron scattering results show that the excited
crystal field levels have energies $E_1$ = 13.2 meV, $E_2$ = 44.8
meV for Ce$_3$In and $E_1$ = 18.5 meV, $E_2$ = 36.1 meV for
Ce$_3$Sn. The Kondo temperature deduced from the quasielastic
linewidth is 17 K for Ce$_3$In and 40 K for Ce$_3$Sn. The low
temperature behavior of the specific heat, magnetization, and
susceptibility can not be well-described  by J=1/2 Kondo physics
alone, but require calculations that include contributions from the
Kondo effect, broadened crystal fields, and ferromagnetic
correlations, all of which are known to be important in these
compounds. We find that in Ce$_3$In the ferromagnetic fluctuation
makes a 10-15 $\%$ contribution to the ground state doublet entropy
and magnetization. The large specific heat coefficient $\gamma$ in
this heavy fermion system thus arises more from the ferromagnetic
correlations than from the Kondo behavior.

\end{abstract}

\vskip 15 pt


\maketitle

\section{Introduction}

In heavy fermion (HF) compounds, it is very common to establish the
Kondo energy scale $T_K$ from the linear coefficient of specific
heat $\gamma$ through Rajan's formula $T_K$= $\pi J R/3 \gamma_0$
derived for the degenerate ($2J+1 \geq$ 2) Kondo model\cite{Rajan}
where $J$ is the total angular momentum. In previous studies of the
specific heat of the HF compounds Ce$_3$X (X=In,
Sn)\cite{YYChen,Lawrence} which crystallize in the Cu$_3$Au cubic
structure, this formula was used to determine the Kondo temperature,
which was found to be 4.8 K for Ce$_3$In and 16.7 K for Ce$_3$Sn.
The crystal electric field (CEF) excitation energy was estimated to
be $T_{CEF}$=65 K.

Most HF compounds reside close to a quantum critical point
(QCP)where antiferromagnetic (AFM) or ferromagnetic (FM)
correlations are present. This makes the previous analysis
inappropriate in so far as it assumes that the magnetic correlations
do not contribute to $\gamma$. Indeed, the Wilson ratios ($\pi^2 R
\chi_0 / 3 C_J \gamma_0$) which were determined previously for
Ce$_3$In and Ce$_3$Sn are 11.5  and 7.0
respectively\cite{YYChen,Lawrence}, indicating that ferromagnetic
correlations dominate the susceptibility.

Inelastic neutron scattering(INS) experiments on single crystals of
compounds that are close to a QCP, such as
CeRu$_2$Si$_2$\cite{CeRu2Si2} or CeNi$_2$Ge$_2$\cite{CeNi2Ge2}
exhibit two classes of excitations. At most $Q$ in the Brillouin
zone, the scattering has the characteristic Kondo energy dependence and is
$Q$-independent or only weakly $Q$-dependent. Similar behavior is
observed in intermediate valence compounds for which it is clear
that the behavior of the low temperature susceptibility, specific
heat and INS spectra are close to the Kondo impurity prediction, as
though the onset of lattice coherence has only a minor effect on
these measurements\cite{YbInCu4,slowcrossover}. Near the QCP,
however, large $Q$-dependent scattering is observed with maximum
intensity at the critical wavevector $Q_c$ ($Q_c$ = 0 for FM and
$Q_c$ = $Q_N$ for AFM) where ordering occurs in the nearby magnetic
state. This scattering represents the short range order. It is
dynamic and critically slows down, or softens, as the QCP is
approached by lowering the temperature or changing a control
parameter. These fluctuations affect the specific heat and can result in non-Fermi
liquid behavior.

Hence INS in single crystals can separate the Kondo behavior from
the contributions due to magnetic correlations. Since the spectral
weight in the magnetic correlations is typically small, INS in
polycrystals will be dominated by the $Q$-independent Kondo
scattering. INS can also be used to directly determine the CEF
excitations. Under these circumstance, INS provides a better way to
determine $T_K$ and $E_{CEF}$ than through analysis of the specific
heat. In this paper, we employ INS to determine both $T_K$ and
$E_{CEF}$. We have re-measured the magnetic susceptibility, and have
extended the specific heat measurement, which in the previous report
was measured down to 1.8 K in zero applied magnetic field, to $T$ =
400 mK and $B =$ 9T. We have also measured the low temperature
magnetization to 13 T.

In the Ce$_3$X compounds, the Ce ions sit on the face centers of the cubic lattice and are subject to
a crystalline electric field (CEF) of tetragonal symmetry. In this case, the Hamiltonian is described as:\\\\
$H_{CF}=B_2^0O_2^0 + B_4^0O_4^0+ B_4^4O_4^4$,\\\\
where $B_l^m$ and $O_l^m$ are the crystal field parameters and Steven operators,
respectively. The sixfold degenerate 4$f^1$, J=5/2 state splits into three
doublets. Diagonalizing the Hamiltonian, the atomic wave functions are
given by:\cite{Aviani,Fischer}\\\\
$\Gamma_7^{(1)}= \eta|\pm 5/2>+ \sqrt {1-\eta^2}| \mp 3/2>$\\
$\Gamma_7^{(2)}= \sqrt{1-\eta^2}|\pm 5/2> - \eta| \mp 3/2>$\\
$\Gamma_6 = | \pm 1/2>$\\\\

Depending on the admixture of the $J$=5/2 and 3/2 states, the inelastic neutron
scattering spectra will exhibit one or two inelastic excitations. Low energy transfer
quasielastic scattering will also be observed if the instrumental resolution is adequate.
From the INS spectra, the crystal field energies and wavefunctions can be determined
from the ampitudes and energies of the excitations. The Kondo effect, which arises from
the hybridization of the 4 $f$-electron with the conduction electrons, broadens the
peak line-widths proportional to $k_BT_K$. The quasielastic scattering peak width
$\Gamma_{QE}$ can be equated to the Kondo energy $k_BT_K$ of
the ground state doublet.

In what follows, we will use the CEF parameters and the Kondo energies derived from the
neutron scattering to calculate the Kondo contribution to the specific heat, susceptibility
and magnetization. All the Kondo calculations utilized\cite{Rajan, Hewson, Sacramento} employ
the same Bethe-Ansatz calculation, making intercomparison possible.

\section{experiment}

All samples were prepared by arc melting in an ultra-high-purity argon atmosphere. After
melting the samples were sealed under vacuum and annealed at 500$^0$C for 2 weeks and cooled
slowly to room temperature. The magnetization was measured in a 14 T Quantum Design Vibrating
Sample Magnetometer at the National High Magnetic Field Laboratory (NHMFL) at Los Alamos National
Laboratory. The specific heat was measured in a Quantum Design PPMS system. The magnetic
susceptibility measurements were performed in a commercial superconducting quantum interference
device (SQUID) magnetometer.

We performed inelastic neutron scattering on a 29 gram sample of
Ce$_3$In and a 37 gram sample of Ce$_3$Sn using the high energy
transfer chopper spectrometer (HET) at ISIS (at the Rutherford
Appleton Laboratory) and the low resolution medium energy chopper
spectrometer (LRMECS) at IPNS (at Argonne National Laboratory). For
Ce$_3$Sn, the quasi-elastic neutron spectrometer (QENS, at IPNS) was
also used to measure the low energy scattering. To increase the
dynamic range of the INS spectrum, a variety of incident energies
($E_i =$ 15 meV, 35 meV, 60 meV, 100 meV for HET and 35 meV for
LRMECS) and temperatures ( 4.5 K, 100 K, 150 K, 200 K and 250 K for
HET; 10K, 100K, 150K for LRMECS) were employed. The HET data have
been normalized to vanadium to establish the absolute value. All the
data have been corrected for absorption ( which is very obvious for
Ce$_3$In case), total scattering cross section, and sample mass.

For the HET data, the low Q data were obtained from averaging the
low angle detectors with angles ranging from 11.5 degrees to 26.5
degrees. The high Q data were obtained from the high angle detector
bank at an angle 136 degrees. For the LRMECS experiment, the low Q
data were obtained by averaging over the low angle detectors with
average angle equal to 13 degrees; and the high Q data were obtained
from high angle detectors where the average angle was 87 degrees.
The QENS data were collected at 7 K. This inverse geometry
spectrometer has 19 detector banks with $Q$ from $Q = 0.36 {\AA}$ to
$Q = 2.52 {\AA}$, each with a slightly different final energy ($E_f$
from 2.82 meV to 3.36 meV). For every fixed $Q$, we removed the Ce
4$f$ form factor to obtain a spectrum representing the $Q=0$
scattering and then summed all 19 spectra together to obtain a total
$S(Q=0, \Delta E)$ spectrum. In order to compare the QENS spectrum
with the spectra from the direct geometry spectrometers HET and
LRMECS, we multiplied the QENS spectrum $S(Q=0, \Delta E)$ by the
4$f$ form factor apppropriate for HET at $E_i$ = 35 meV.

To subtract the nonmagnetic (background, single phonon, and multiple phonon) contributions,
we measured the non-magnetic counterpart compounds La$_3$In
and La$_3$Sn. For the specific heat, we obtained the magnetic contribution by direct
subtraction, i.e. $C_{mag}= C(Ce)-C(La)$. For the INS data we
used La$_3$In and La$_3$Sn to determine the scaling of the nonmagnetic scattering
between low $Q$ and high $Q$ as $h(\Delta E)=S(La,LQ)/S(La,HQ)$. Using this factor we
scaled the high $Q$ data (where the nonmagnetic scattering dominates) to the low $Q$ data
(where the magnetic scattering dominates) in Ce compounds to determine the nonmagnetic
contribution\cite{Murani,Eugene,Jon}.

\section{results and discussion}

Fig. 1(a) and (c) directly compare the low $Q$ INS spectra of Ce$_3$In and
La$_3$In; the data were collected on HET with incident energy $E_i=$ 15 meV (a)
and 60 meV (c) at 4.5 K. Fig. 1(b) compares the low $Q$ and high $Q$ data for
Ce$_3$In collected on LRMECS with an incident energy $E_i$= 35 meV at 10 K. The
low $Q$ data for Ce$_3$Sn and La$_3$Sn, which were collected on HET, are compared
in Fig. (d), (e) and (f) where the incident energies are $E_i$=15 meV (d), 35 meV (e)
and 60 meV (f) at 4.5 K. In these spectra, two excited energy levels, corresponding to
crystal field excitations, are observed for both Ce$_3$In and Ce$_3$Sn. The
spectra (a) and (d), which compare the
Ce$_3$In(Sn) and La$_3$In(Sn) scattering at low energy transfer ($\Delta E<$ 9 meV),
exhibit obvious quasielastic scattering which as mentioned above arises from
Kondo scattering.

\begin{figure}[t]
\centering
\includegraphics[width=0.5\textwidth]{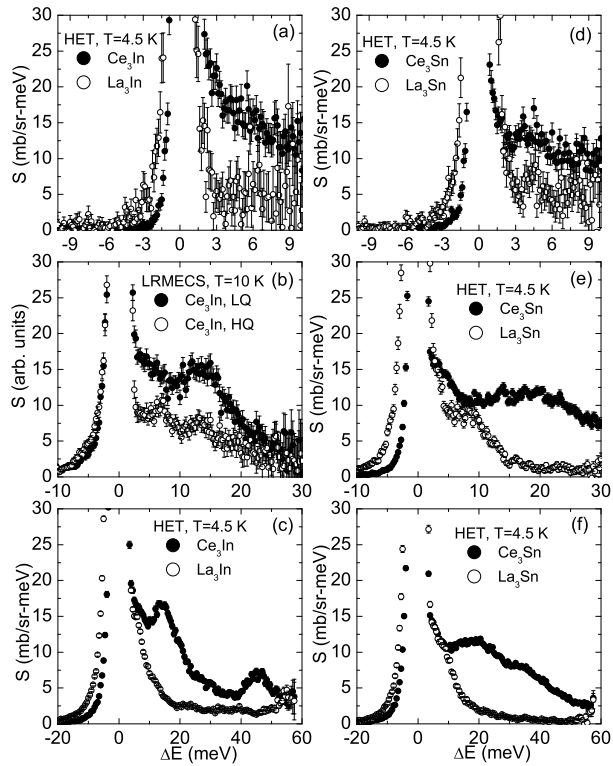}
\caption{\label{fig:1} Inelastic neutron scattering spectra for
Ce$_3$In and Ce$_3$Sn together with that of their nonmagnetic
counterpart compounds obtained from HET and LRMECS. The data
collected on HET are at 4.5 K and on LRMECS are at 10 K. (a)
$E_i$=15 meV, (b) $E_i$=35 meV, and (c) $E_i$=60 meV spectra of
Ce$_3$In and La$_3$In. (d) $E_i$=15 meV, (e) $E_i$=35 meV and (f)
$E_i$=60 meV spectra for Ce$_3$Sn and La$_3$Sn. All spectra are for
low $Q$ except in (b) where a high-$Q$ spectrum is included for
comparison. }\vspace*{-3.5mm}
\end{figure}

The magnetic contribution $S_{mag}$ to the scattering of Ce$_3$In,
obtained using the method described above, is shown in Fig. 2.  The
solid lines represent a fit to the CEF model. Since the inelastic
peaks are relatively broad, the line widths $\Gamma_i$ are taken to
be finite. In this case,
the magnetic scattering is described as:\\\\
$S_{mag} =\frac{2N}{\pi \mu_B^2} f^2(Q)(1-e^{-\Delta E/{k_B T} }) \chi^{\prime\prime}(Q,\Delta E)$\\\\
$\chi^{\prime\prime}(Q,\Delta E) = \Sigma \chi_i(T) \Delta E (\frac{\Gamma_i}{2\pi})/[(\Delta E - E_i)^2 + \Gamma^2_i]$\\\\
Here $i$=0,1,2, $E_0$ = 0 corresponds to the quasielastic
scattering, and $f^2(Q)$ is the Ce 4$f$ form factor. The CEF model
fitting was performed simultaneously on six data sets at three
different incident energies ($E_i$=15 meV, 35 meV and 60 meV) and at
two different temperatures (4.5 K and 150 K). Fig. 2(a)-(d) are the
data collected on HET. In Fig. 2(f) the LRMECS data are displayed
for comparison. The resulting CEF fitting parameters are shown in
Table I. The ground state is the $\Gamma_7^{(1)}$ doublet, the first
excited state is the $\Gamma_7^{(2)}$ doublet\cite{endnote1} at the
energy 13.2 meV, and the second excited state is the $\Gamma_6$
doublet at the energy 44.8 meV. The quasielastic line width
$\Gamma_{QE}$ = $\Gamma_0$ = 1.49 meV, implies that $T_K$ =
$\Gamma_{QE}/k_B$ = 17 K.

\begin{table}[htp]
\caption{\label{tab:table} CEF model fitting parameters for Ce$_3$In and Ce$_3$Sn.}
\begin{ruledtabular}
\begin{tabular}[b]{ccc}
       &Ce$_3$In  & Ce$_3$Sn \\
\hline
$B_2^0$(meV) & -2.203$\pm$0.015 & -1.660$\pm$0.017 \\
$B_4^0$(meV) &  0.066$\pm$0.001 & 0.038$\pm$0.0009 \\
$B_4^4$(meV) &  -0.154$\pm$0.004 & -0.263$\pm$0.003 \\
$\eta$       & 0.94 &  0.89  \\
$E_1 $ (meV) & 13.2 & 18.5  \\
$E_2 $ (meV)  & 44.8 & 36.1  \\
$\Gamma_{QE}$ (meV) & 1.49$\pm$0.07 & 3.52$\pm$0.16  \\
$\Gamma_1$ (meV) & 5.98$\pm$0.07 & 9.37$\pm$0.038  \\
$\Gamma_2$ (meV) & 2.06$\pm$0.37 & 6.28$\pm$0.38  \\
$\chi^2$  & 2.4057 &  2.1528  \\
$\lambda$ (mole-Ce/emu) & 62 & 85 \\

\end{tabular}
\vspace{-2mm}
\end{ruledtabular}
\end{table}

\begin{figure}[t]
\centering
\includegraphics[width=0.5\textwidth]{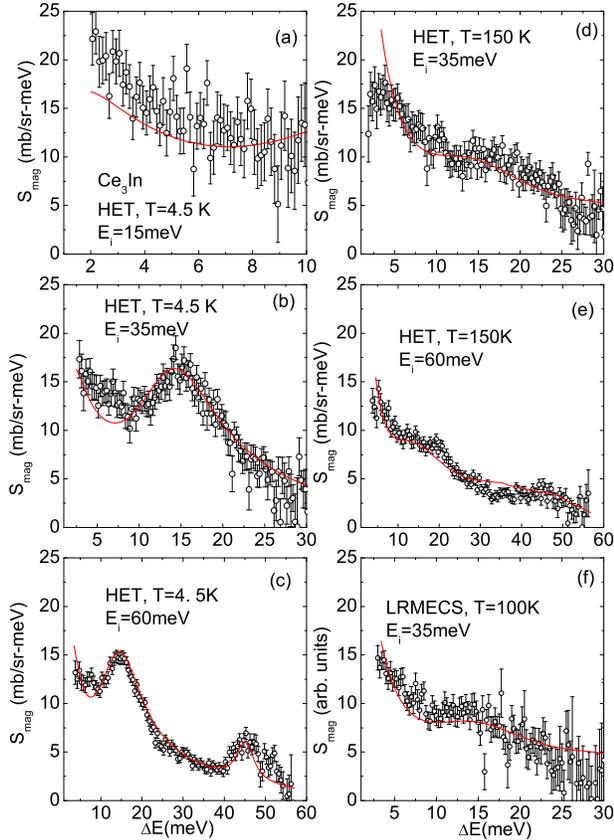}
\caption{\label{fig:2} Magnetic contribution $S_{mag}$ to the
inelastic neutron scattering spectra of Ce$_3$In for data taken on
HET at T=4.5 K and 150 K with different incident energies $E_i$=15
meV, 35 meV, and 60 meV and taken on LRMECS at T=100 K with $E_i$=35
meV. The solid lines represent the quasielastic and crystal field
contributions obtained from least squares fitting as described in
the text. } \vspace*{-3.5mm}
\end{figure}

In Fig. 3(a)-(e) we display the magnetic contribution to the Ce$_3$Sn scattering collected
from HET at 4.5 and 100 K and at three incident energies (15, 35, and 60 meV). Data from QENS
at 7 K (Fig. 3(f)) are included for comparison. The CEF fits are also included (solid lines);
as for the Ce$_3$In case, the fits were performed simultaneously on six different spectra at
different incident energies and temperatures. The intensity and form factor of the QENS data
have been adjusted to that of the HET spectra at $E_i$=35 meV (spectra (b)) to make a direct
comparison. The fitting parameters yield a similar crystal field scheme as for Ce$_3$In:
the $\Gamma_7^{(1)}$ doublet is the ground state, $\Gamma_7^{(2)}$ is the first excited
state with energy 18.5 meV, and the second excited state is the $\Gamma_6$ doublet  at the
energy 36.1 meV. The Kondo temperature $T_K =$ 40 K is higher than for Ce$_3$In, and the
excited state linewidths are broader, reflecting stronger 4$f$-conduction hybridization.

\begin{figure}[t]
\centering
\includegraphics[width=0.5\textwidth]{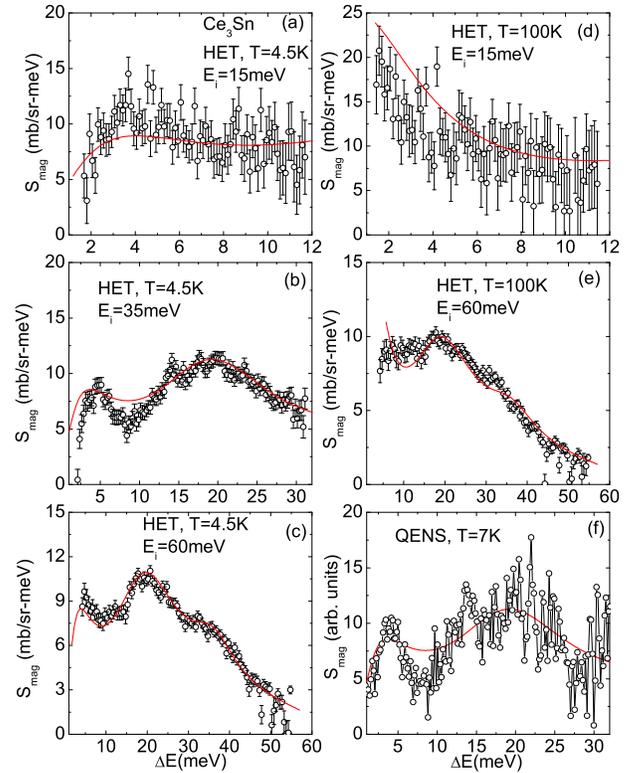}
\caption{\label{fig:3} (a)-(f) Magnetic contribution $S_{mag}$ to the inelastic neutron
scattering spectra for Ce$_3$Sn. The temperatures and incident energies are given in
the plot. The solid lines represent the CEF model.  (f): the magnetic contribution to
the INS spectra collected from QENS at 7 K.  The solid line in (f) is the CEF model fit
for the $E_i$ = 35 meV spectra.
} \vspace*{-3.5mm}
\end{figure}

Due to the large CEF excitation energies, the low temperature
behavior of the magnetic specific heat should be dominated by the
$\Gamma_7^{(1)}$ doublet ground state. This is confirmed by the fact
that the magnetic entropy (Fig. 4(a) inset) reaches Rln2 near 20 K
but only reaches Rln4 near 70 K. For a doublet ground state, the
Kondo model predicts $\gamma_0 = \pi R / 6 T_K$ for the linear
coefficient of specific heat \cite{Rajan}. In previous results for
Ce$_3$In a value $T_K =$ 4.8 K was deduced using this
formula\cite{YYChen}. In addition, the specific heat coefficient
$C/T$ showed a peak near 2 K  whose existence was somewhat uncertain
since the lowest measured temperature was only 1.8 K. We have
extended the specific heat measurement down to 400 mK. In Fig. 4(a)
we plot $C_{mag}/T$ and find a peak at $T$ = 2.6 K. Comparison of
the data to the prediction $\gamma^{K}(T)$ of the Kondo model which
is calculated using the value $T_K = \Gamma_{QE} / k_B = 17 K$
deduced from our neutron data, shows that the Kondo prediction is
much smaller than the experimental value; indeed, $\gamma^{K}_0$ is
only half of $\gamma^{exp}_{0.4 K}$ (Table II). Given the large
Wilson ratio reported earlier\cite{YYChen}, the obvious explanation
is that ferromagnetic (FM) fluctuations dominate the low temperature
specific heat, increasing the specific heat above the Kondo value
and giving rise to the peak at 2.6 K representing the onset of short
range FM order.

We next consider the high temperature susceptibility, comparing the
measured value to the value calculated from the crystal field
parameters of Table I in the inset of Fig. 4(b). A molecular-field
$\lambda$ = 62 mole-Ce/emu has been added to compensate the
reduction of the susceptibility at high temperature due to the Kondo
effect ($1/\chi^{HT}=1/\chi^{CEF}+\lambda$). At high temperatures,
when the crystal field states are excited, the effective Kondo
temperature $T^{HT}_K$ is larger than the Kondo temperature of the
ground state doublet. The molecular field constant is related to the
effective Kondo temperature via $\lambda = T^{HT}_K / C_{5/2}$ where
$C_{5/2}$ is the free ion Curie constant for cerium. This relation
gives  $T^{HT}_K =$ 77 K, which value is essentially equal to the
width $\Gamma _1$ of the first excited level seen in the neutron
scattering (Fig. 2 and table I).

At low temperatures, there should be three contributions to
$\chi(T)$, as well as to $M(H)$ and $C_{mag}$: one from the Kondo
single ion impurity physics of the ground state doublet, one from
the FM fluctuations, and one from the excitation of higher lying
crystal field states. To carry out such an analysis, we note first
that in the Cu$_3$Au crystal structure, the tetragonal crystal field
axis (i.e. the z-axis for the doublet wave functions) points
perpendicular to the face containing any given face-centered cerium
atom; hence there are three orthogonal tetragonal axes in the unit
cell. When applying a magnetic field in a polycrystalline sample,
the field will point along the tetragonal axis for $\frac{1}{3}$ of
the cerium atoms but orthogonal to the tetragonal axis (in the $x-y$
plane) for $\frac{2}{3}$ of the atoms. The effective low temperature
Curie constant is then $C_{eff}= \frac{1}{3} C^z_{eff}+ \frac{2}{3}
C^x_{eff}$, where $C^{z(x)}_{eff}= N (g^{z(x)}_{eff} \mu_B)^2
\frac{1}{2}(\frac{1}{2}+1)/ 3k_B$. This is the form for a pseudo
spin $\frac{1}{2}$ doublet where the CEF physics is absorbed into
the effective $g$-factor. Here $g^{z(x)}_{eff} = \frac{12}{7}
<J_{z(x)}>$ where $<J_{z(x)}>$ is the matrix element of the angular
moment component along the $z(x)$ axis. From the CEF mixing
parameter $\eta$, we determine $C_{eff}$ to be 0.48 emu-K/mole-Ce
for Ce$_3$In and 0.41 emu-K/mole-Ce for Ce$_3$Sn. (Table II).

\begin{table*}[htp]
\caption{\label{tab:table} Kondo single ion model calculation for Ce$_3$In and Ce$_3$Sn.}
\begin{ruledtabular}
\begin{tabular}[b]{cccccccccc}
  & $<J_z>$ & $<J_x>$ & $M^{sat}_{CEF}$ &  $C^{LT}_{eff}$ &  $\chi^K_0(\frac{emu}{mole-Ce})$ & $\gamma^K_0(\frac{J}{mole Ce K^2})$ &  $\chi^{exp}_{0.4K}(\frac{emu}{mole-Ce})$ & $\gamma^{exp}_{0.4K}(\frac{J}{mole Ce K^2})$ \\
\hline
Ce$_3$In & 2.0344 & 0.7171 & 0.991 &  0.4765 & 0.0076  & 0.256 & 0.064 & 0.467\\
Ce$_3$Sn & 1.6684 & 0.9074 & 0.994 &  0.4086 & 0.0033 & 0.109 &0.018 &  0.221\\

\end{tabular}
\vspace*{-2mm}
\end{ruledtabular}
\end{table*}

\begin{figure}[t]
\centering
\includegraphics[width=0.45\textwidth]{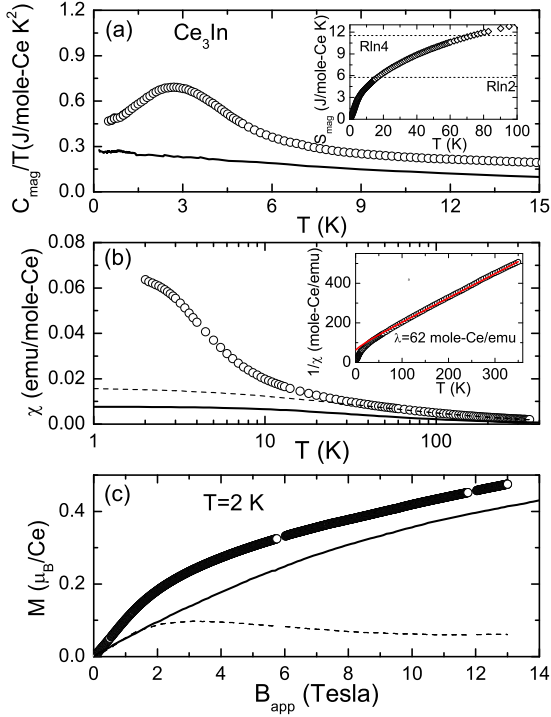}
\caption{\label{fig:4} (a) The magnetic contribution to the specific
heat $C_{mag}/T$ versus $T$ for Ce$_3$In. The solid line is the
Kondo prediction $\gamma^{K} (T)$ calculated using $T_K =
\Gamma_{QE}/k_B =$ 17 K. The inset is the magnetic entropy of
Ce$_3$In. (b) The magnetic susceptibility $\chi (T)$ for Ce$_3$In.
The solid line is the Kondo prediction $\chi^{K}(T)$ calculated
using $T_K =$ 17 K and the low temperature Curie constant determined
as described in the text. The dashed line gives the sum of the Kondo
and crystal field contributions $\chi^{CEF+K}=\chi^{K}+ (\chi^{CEF}-
\chi^{LT}_{Curie})$. The inset is the inverse susceptibility
together with the calculated susceptibility (solid line)
$1/\chi^{HT}=1/\chi^{CEF}+\lambda$ obtained using the CEF fitting
parameters in table I. (c) The magnetization for Ce$_3$In. The solid
line $M^{K}(B)$ is the Kondo calculation calculation using $T_K = $
17 K. The dashed line is the contribution from the ferromagnetic
fluctuations. } \vspace*{-3.5mm}
\end{figure}

To sort the low temperature susceptibility into Kondo, FM, and CEF
contributions, we note that since the first CEF excited level is at
152 K, at sufficiently low temperatures the Zeeman splitting of the
$\Gamma_7^{(1)}$ ground state doublet will obey a Curie law
$\chi^{LT}_{Curie} = C^{LT}_{eff} / T$. Due to the Kondo effect,
this Curie behavior will be replaced by the Kondo contribution
$\chi^{K}$, which we calculate using the same Curie constant
$C^{LT}_{eff}$ and using $T_K =$ 17 K (solid line, Fig. 4(b)). The
susceptibility from the combination of the ground state Kondo and
the excited crystal fields will then be of the form
$\chi^{CEF+K}=\chi^{K}+ (\chi^{CEF}- \chi^{LT}_{Curie})$ where we
subtract $\chi^{LT}_{Curie}$ to avoid double counting the ground
state contribution. As for the specific heat, the resulting
$\chi^{CEF+K}$ (dashed line in Fig. 4(b)) is much smaller than the
experimental value at $T < $ 20 K. The excess can be viewed as the
contribution from the ferromagnetic fluctuations. Taking the latter
to be equal to the difference $\chi^{exp} - \chi^{CEF+K}(T)$, the FM
contribution is seen to increase below 10 K in a manner
characteristic of ferromagnetic short range order.

In Fig. 4(c) we exhibit the magnetization as measured up to 13 Tesla
at $T$=2 K. Based on Hewson's calculation of Kondo
model\cite{Hewson}, we can estimate the Kondo contribution to the
magnetization. Since the effective $g$-factors differ in the $z$ and
$x-y$ directions, we calculate $M^{K} = \frac{1}{3} M^{K} (z) +
\frac{2}{3} M^{K}(xy)$. The result is plotted as a solid line in
Fig. 4(c). After subtracting the Kondo contribution, we obtain the
contribution from the FM correlations (dashed line). This saturates
at a relatively small field $B \sim 2.5$ tesla with $M^{sat}$ =
0.095 $\mu_B$, which is 10 percent of the saturation value 1.0
$\mu_B$ expected based on the effective $g$-factors.

\begin{figure}[t]
\centering
\includegraphics[width=0.45\textwidth]{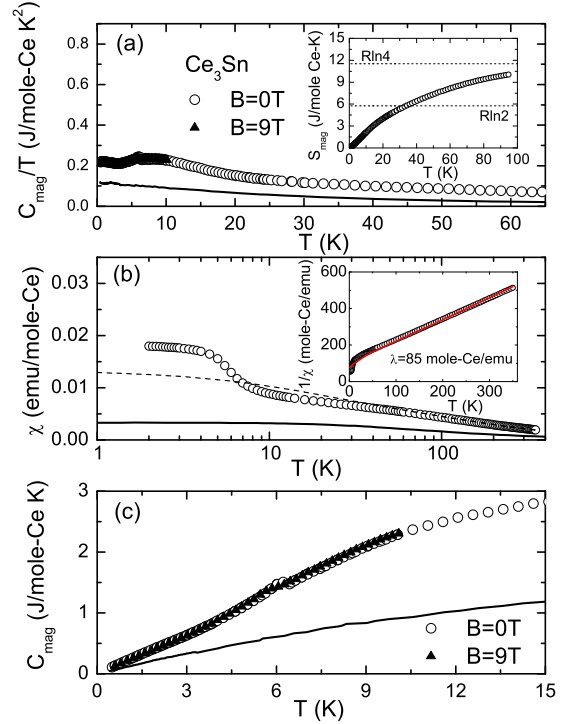}
\caption{\label{fig:5} (a) the magnetic specific heat $C_{mag}/T$
versus $T$ at B=0 T (open circle) and B=9 T (solid triangle) for
Ce$_3$Sn. The solid line is the Kondo contribution $\gamma^{K} (T)$
for $T_K =$ 40 K. The inset is the magnetic entropy. (b) Magnetic
susceptibility $\chi (T)$ for Ce$_3$Sn. The solid line is the Kondo
contribution $\chi^{K}(T)$ calculated with $T_K = \Gamma_{QE}/k_B =$
40 K and the dashed line is the sum of the Kondo and CEF
contributions $\chi^{CEF+K}=\chi^{K}+ (\chi^{CEF}-
\chi^{LT}_{Curie})$. The inset is the inverse susceptibility
together with the value $1/\chi^{HT}=1/\chi^{CEF}+\lambda$ (solid
line) calculated from CEF fitting parameters in table I.  (c)
$C_{mag}$ at $B$=0 T (open circle) and B=9 T (solid triangle) for
Ce$_3$Sn. The thin solid line is $C^{K}(T)$ calculated with $T_K =$
40 K.} \vspace*{-3.5mm}
\end{figure}

In the same way as for Ce$_3$In, we calculate $\chi^{K}(T)$,
$\chi^{CEF}(T)$, $\chi^{CEF+K}(T)$, $\gamma^{K}(T)$ and $C^{K}(T)$
for Ce$_3$Sn, comparing to the measured data in Fig. 5. The high
temperature susceptibility (Fig. 5(b) inset) can again be fit with
the sum of the CEF contribution calculated using the parameters of
Table I and a molecular field contribution (solid line). The value
$\lambda =$ 85 mole-Ce/emu of molecular field constant implies an
effective Kondo temperature at high temperature $T^{HT}_K$ = 105 K,
which again is essentially equal to the linewidth 9.4 meV of the
first excited state seen in the neutron scattering (Table I).

The solid lines in Fig. 5 represent the Kondo ground state doublet
contributions. In Fig. 5(b), the dashed line is $\chi^{CEF+K}(T)$.
The excess due to the FM correlations has a much smaller magnitude
($\sim$ 0.005 emu/mol-Ce) than for Ce$_3$In where the FM
contribution is of order 0.05 emu/mol-Ce. A similar statement holds
for the FM contribution to the specific heat coefficient, which is
of order 0.4 J/mol-Ce-K$^2$ for Ce$_3$In but only 0.1 J/mol-Ce-K$^2$
for Ce$_3$Sn (Figs. 4(a) and 5(b)). Hence the FM enhancement is
smaller in Ce$_3$Sn than in Ce$_3$In, consistent with the larger
value of $T_K$.

In order to better understand these compounds, we measured the
specific heat of Ce$_3$In under different applied fields (B = 0 T, 1
T, 3 T, 6 T and 9 T). The results for the magnetic contribution
$C_{mag}$ are shown in Fig. 6. The low temperature peak in $C_{mag}$
moves to higher temperature when the field is increased. Since the
peak in the Kondo contribution to $C_{mag}$ is expected to increase
with field, we plot in Fig. 6(a)-(e) the Kondo contribution $C^{K}
(B)$ calculated for different applied fields using the theoretical
results of Sacramento and Schlottmann\cite{Sacramento}. In
calculating $C^{K} (B)$, we again account for the different
effective $g$-factors in the $z$ and $x-y$ directions. The results
indicate that the Kondo contribution is not expected to alter
significantly in applied fields of order 9 T, essentially because
$g_{eff}\mu_B B < k_B T_K$ for these fields. This makes it clear
that the peak does not arise from the Kondo scattering but must be
due to the FM  fluctuations.

To quantify the FM contribution, we again assume that the measured
magnetic specific heat is the sum of the ground state doublet Kondo
contribution $C^{K} (B)$, the FM contribution $C^{FM}$, and a
contribution $C^{CEF}$ due to the excitation of higher lying CEF
states. Since the FM fluctuations appear to only contribute to the
susceptibility below 10 K (Fig. 4 b) we assume that the excess
$C_{mag}$ - $C^{K} (B)$ observed for $T>$ 10 K is primarily due to
CEF excitations. Given the large linewidths of the CEF excitations
seen in the neutron scattering, and concomitant large effective
$T^{HT}_{K}$ at high temperature, this contribution to the specific
heat is much broader as a function of temperature than would be the
case for a simple CEF Schottky anomaly. For simplicity, we
approximate this CEF contribution as linear in temperature, with
slope equal to that observed in the range 8-15 K, and we assume that
since the CEF excitation energy is large, this contribution will be
unaffected by fields of order 9 T. We approximate the FM
contribution $C^{FM}$ as a Gaussian, centered at a temperature that
increases with field. The three contributions, Kondo, CEF, and FM,
are plotted at the different fields in Figs. 6(a)-(e). The solid
lines, which represent the sum of all three contributions, fit the
data very well at all fields.

We plot the Gaussian peak temperature in Fig. 6(f), where it is seen
to grow linearly with field. This suggests Zeeman splitting, where
at zero field the splitting arises from the internal field in the
regions of FM short range order, and where the applied field
increases the splitting. To determine the internal field $B_{int}$,
we calculate the Schottky anomaly $C^{Schotkky} (B_{int})$ expected
due to Zeeman splitting of a doublet with the same effective
$g$-factors as we have obtained from the neutron fits; we then
adjust $B_{int}$ until the peak temperature of the Schottky anomaly
is the same as that of the Gaussian peak temperature for $B$ = 0.
This gives $B_{int}$ = 9.5 T. We then calculate $C^{Schottky}
(B_{int} + B_{app})$ to determine peak position of the Schottky
anomaly in an applied field $B_{app}$. As can be seen in Fig. 6 (f),
the Gaussian peak temperatures track the expected Zeeman splitting
very closely. On the other hand, the temperature dependence of the
Schottky specific heat calculated in this manner is considerably
broader than the Gaussian contributions $C^{FM}$ that are plotted in
Fig. 5. This means that, while the contribution of the FM short
range order to the specific heat is not of Schottky form, the
increase of the Gaussian peak position is the same as the Zeeman
splitting expected for a total field $B_{int}+B_{app}$ given the
effective $g$-factors.

The entropy of the Gaussian contribution is about 15 $\%$ of Rln2
for all fields. This corresponds to the estimate obtained from the
magnetization $M(B)$ where the saturation value of the FM
contribution is about 10$\%$ of the value 1.0 $\mu_B$ expected for
the $\Gamma_7^{(1)}$ ground state doublet for the measured value of
$\eta$. Hence, the enhancement of $\chi_0$ and $\gamma_0$ arises
from magnetic fluctuations which involve 10-15 $\%$ of the 4$f$
electron degrees of freedom.

In Fig. 5(c), we compare the magnetic specific heat $C_{mag}$ at
zero field and $B=$ 9 T for Ce$_3$Sn. The solid line is the Kondo
contribution $C^{K}$. The specific heat does not change with field
for $B<$ 9 T. The most likely explanation of this is that, as
discussed above, the FM correlations make a smaller contribution
than in Ce$_3$In. The excess specific heat $C_{mag} - C^{K}$ seen
for $T >$ 6 K is presumably due to the CEF contribution, which
should be even broader in temperature in Ce$_3$Sn than in Ce$_3$In
due to the larger Kondo temperature.

\begin{figure}[t]
\centering
\includegraphics[width=0.5\textwidth]{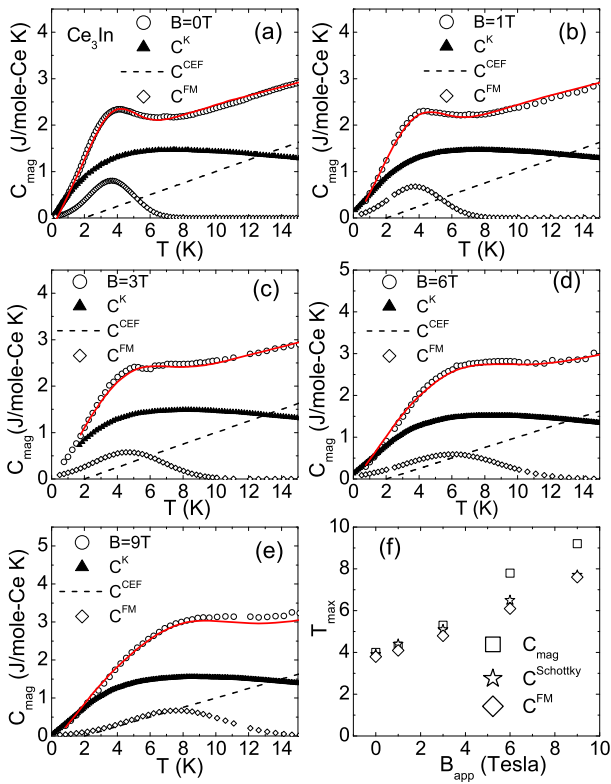}
\caption{\label{fig:6} (a), (b), (c), (d) and (e): $C_{mag}$ of Ce$_3$In in different
applied magnetic fields. The solid line sums the three contributions (Kondo, CEF, and
FM fluctuations) shown in the plot.
(f): The peak position of $C_{mag}$, of the Gaussian contribution $C^{FM}$ due to
the FM fluctuations, and the expected peak position in the Schottky anomaly $C^{Schottky}(B_{int}+B_{app})$
due to Zeeman splitting in the presence of an internal field $B_{int}$.
}\vspace*{-3.5mm}
\end{figure}

We have demonstrated that the large $\gamma$ observed in Ce$_3$In
arises more from ferromagnetic correlations than from the single ion
Kondo physics. This reflects the fact that the system is close to a
ferromagnetic quantum critical point. In a $Q$-resolved INS
experiment, the ferromagnetic correlations should show up in the
vicinity of $Q=$0 riding on a background of $Q$ independent Kondo
scattering. We can estimate that these FM correlations will have
10-15$\%$ of the total spectral weight in $Q$-space. Since the large
$\gamma$ and the proximity to the QCP occurs when the Kondo
temperature $T_{K}=$ 17 K is fairly large,  it is also reasonable to
believe that when an appropriate control parameter (e.g. alloying
parameter x in Ce$_{3-x}$La$_x$In) drives this system to the QCP,
the Kondo temperature $T_{K}$ will remain finite, as expected for
example for a  spin density wave type QCP.

We have observed strong FM fluctuations in the related compound
Pr$_3$In, which is an antiferromagnet below 12 K\cite{Andy}. A
possibility for this behavior is that AFM interactions between rare
earth atoms on the face centers of the Cu$_3$Au structure are
frustrated. If, for example, the atoms at (1/2 1/2 0) and (1/2 0
1/2) are aligned antiferromagnetically, the atom at (0 1/2 1/2) will
be free to point to any direction. Ferromagnetic
next-nearest-neighbor interactions could then stabilize
ferromagnetism on this sublattice\cite{Cristian}. In any case, the
FM correlations appear to be generic to this crystal structure.

In conclusion, we have used inelastic neutron scattering to
determine the crystalline electric field (CEF) splitting and Kondo
energy scale in Ce$_3$In and Ce$_3$Sn. For both compounds the
crystal field excitation energy is large. For Ce$_3$In we have
separated the magnetization $M(B)$, susceptibility $\chi(T)$ and
specific heat $C_{mag}$ into contributions from the Kondo effect,
from the CEF, and from FM fluctuations. The simplified model
calculation for Ce$_3$In shows that the FM correlations make a 15
$\%$ contribution to the doublet ground state entropy and that the
large $\gamma$ arises mostly from the FM correlations. This suggests
Ce$_3$In is close to a quantum critical point (QCP). The Kondo
temperature  $T_{K}$ is expected to remain finite at the QCP, as
occurs for a spin density wave type QCP. INS experiments in single
crystals of these compounds would be very interesting.

\section{acknowledgements}

We thank Vivien Zapf for her assistance in the measurement at NHMFL
and Cristian Batista for his insightful comments. Research at UC
Irvine was supported by the U.S. Department of Energy, Office of
Basic Energy Sciences, Division of Materials Sciences and
Engineering under Award DE-FG02-03ER46036. Work at ORNL was
supported by the Scientific User Facilities Division Office of Basic
Energy Sciences, DOE and was managed by UT-Battelle, LLC, for DOE
under Contract DE-AC05-00OR22725. Work at Los Alamos National
Laboratory was performed under the auspices of the U.S. DOE/Office
of Science. Work at NHMFL-PFF, Los Alamos was performed under the
auspices of the National Science Foundation, the State of Florida,
and U.S. DOE. Work at ANL was supported by DOE-BES under contract
DE-AC02-06CH11357.

\end{document}